%% file: v2d.tex
\documentclass[a4paper,fleqn,usenatbib,article]{mnras}
\usepackage{graphicx}
\usepackage{amssymb}
\usepackage{amsmath}
\usepackage[T1]{fontenc}
\usepackage{ae,aecompl}
\usepackage{natbib,times}
\usepackage{lscape}
\usepackage{pdflscape}

\begin{document}
\title[Decompose temporal variations of pulsar DMs]
      {Decompose temporal variations of pulsar dispersion measures}

\author[P. F. Wang and J. L. Han]
       {P. F. Wang$^{1,3}$\thanks{E-mail: pfwang@nao.cas.cn}
         and J. L. Han$^{1,2,3}$\thanks{E-mail: hjl@nao.cas.cn}\\
1. National Astronomical Observatories, Chinese Academy of
Sciences, A20 Datun Road, Chaoyang District, Beijing 100101, China \\
2. School of Astronomy and Space Sciences, University of the
Chinese Academy of Sciences, Beijing 100049, China \\
3. CAS Key Laboratory of FAST, NAOC, Chinese Academy of Sciences, Beijing 100101, China
}

\date{Accepted XXX. Received YYY; in original form ZZZ}

\label{firstpage}
\pagerange{\pageref{firstpage}--\pageref{lastpage}}
\maketitle

\begin{abstract}
  Pulsar dispersion measure (DM) accounts for the total electron
  content between a pulsar and us. High-precision observations for
  projects of pulsar timing arrays show temporal DM variations of
  millisecond pulsars. The aim of this paper is to decompose the DM
  variations of 30 millisecond pulsars by using Hilbert-Huang
  Transform (HHT) method, so that we can determine the general DM
  trends from interstellar clouds and the annual DM variation curves
  from solar wind, interplanetary medium and/or ionosphere. We find
  that the decomposed annual variation curves of 22 pulsars exhibit
  quasi-sinusoidal, one component and double components features of
  different origins. The amplitudes and phases of the curve peaks are
  related to ecliptic latitude and longitude of pulsars, respectively.
\end{abstract}



\section{INTRODUCTION}

Pulsar signals propagate through the ionized medium between a pulsar and
us, and suffer from an extra dispersive delay from the medium as being
\begin{equation}
  t_{\nu}=\frac{e^2}{4\pi m_{\rm e}c} \frac{\int_{\rm us}^{\rm
      psr}{n_{\rm e}(l) {\rm d}l}}{\nu^2},
  \label{eq:dt}
\end{equation}
depending on the frequency of signals $\nu$. Here $c$ is the speed of
light, $m_{\rm e}$ and $e$ are the mass and charge of electrons,
$n_{\rm e}(l)$ represents the number density of electrons along the
sight line, and ${\rm d}l$ is the element distance. In pulsar
astronomy, dispersion measure (DM, in units of cm$^{-3}$pc) is defined
to account for the total electron column density between a pulsar
and us, as being
\begin{equation}
  {\rm DM}=\int_{\rm us}^{\rm psr}{n_{\rm e}(l) dl}.
  \label{eq:dm}
\end{equation}
The observed DM of a pulsar includes the contributions from the
ionized interstellar medium, the inter-planetary medium in the Solar
system and the ionosphere around the Earth. For DMs of most pulsars,
the ionized interstellar medium is predominant, and the contributions
from the ionosphere and the inter-planetary medium are often negligible
but known to affect the DMs of a few pulsars located at low ecliptic
latitudes in some seasons \citep[e.g.][]{yhc+07}.

Small amplitude DM variations have been observed for some pulsars
\citep[e.g.][]{lps88, pkj+13}, which may reflect the drifting of
interstellar clouds into or away from our line of sight to a pulsar or
the changes of electron density distribution in interplanetary medium
or even in the ionosphere. Recent observations with wide-band
receivers or quasi-simultaneous multiple-band observations can
determine pulsar DMs accurately \citep[e.g.][]{kcs+13, rhc+16}, even
up to an accuracy of $10^{-4}$ cm$^{-3}$pc depending on the steepness
of pulse profiles and the frequency range of observations. Temporal DM
variations, if they are not well discounted, would affect the high
precision timing of millisecond pulsars
\citep[e.g.][]{kcs+13,lbj+14,abb+18}, which is therefore one of the
primary sources of low-frequency ``noise'' for measurements of pulse
times of arrival. There have been lots of efforts to eliminate the
``DM noise'' in the pursuit of gravitational wave detection, as done
for Parkes Pulsar Timing Array \citep{kcs+13,rhc+16}, European Pulsar
Timing Array \citep{cll+16,dcl+16}, North American Nanohertz
Observatory for Gravitational Waves \citep{dfg+13,abb+15,abb+18}, and
their combination for the International Pulsar Timing Array
\citep{lsc+16}.

Through Bayesian methodology \citep[e.g.][]{abb+15,lsc+16}, DM
variations can be analyzed for yearly DM variations, non-stationary DM
events and spherically symmetric solar wind term
\citep[e.g.][]{lsc+16}, or simply be represented by a series of
discrete DM values at each epoch \citep[e.g.][]{abb+15,abb+18}. A
number of methods have been developed to explore the temporal DM
features and their power spectrum. For example, linear and periodic
functions and their combinations have been fitted to the DM time
series to get the temporal scales and the trends for DM variations
\citep{jml+17}. Power spectral analyses of DM time series have been
conducted to search for periodic DM modulations
\citep[e.g.][]{kcs+13}. Structure functions were employed to estimate
the power for the stochastic, white noise and periodic components of
the DM time series, and to check the Kolmogorov feature of the
interstellar turbulence
\citep[e.g.][]{yhc+07,kcs+13,rhc+16,lcc+16,jml+17}. The trajectories
of sight lines to pulsars sometimes were plotted to help to interpret
the trends and annual variations \citep{kcs+13,jml+17}. To understand
the origins of DM variations, \citet{lcc+16} carried out a detailed
modeling. They attributed the linear variations to the persistent
gradient of interstellar medium transverse to the lines of sight
and/or the parallel motion between the pulsar and observer. Periodic
variations were attributed to the combined effects of Earth's annual
motion and DMs contributed by the solar wind \citep{yhc+07b,odz+15}
and also heliosphere and plasma lens in the interstellar medium
\citep{lcc+16}. These variations are generally have a period of one
year, while the periodicity induced by ionosphere might be semi-annual
\citep{hr06}. Stochastic DM variations are generally attributed to the
interstellar turbulence, maybe in the form of Kolmogorov turbulence.

The long-term DM variations, often talked as the DM trends, have to be
decomposed from the entire data series. Though the linear trends were
often fitted to data, the real DM variations generally exhibit much
more complicated structures rather than monotonic decreasing or
increasing \citep[e.g.][]{abb+15, abb+18}. When the DM trend cannot be
well decomposed, the properties and relative contributions of the
trend, periodic and stochastic DM variation components would remain
unclear which would be a barrier to understand their origins. The
trends of the complicated DM variations have been analyzed by cutting
data into discrete pieces, and then fitting each section by the
combined triangle function and linear term \citep[e.g.][]{jml+17}.

The DM variations caused by the inter-planetary medium should be
correlated among the pulsars \citep{lcc+16}, depending on their
ecliptic latitudes and longitudes, while DM variations resulting from
the interstellar medium should be uncorrelated. Previous efforts have
been focused on DMs of individual pulsar, and no joint analyses of DM
variations of an ensemble of pulsars have ever been done as we do in
this paper.

In this paper, we employ the Hilbert Huang Transform
\citep[HHT,][]{hsl+98,hsl99, wh09} method to decompose temporal DM
variations of pulsars into components from different physical
origins. This recently developed signal processing method provides a
new tool to decompose different contributions of data series. Since
this is the first time to have the HHT applied in pulsar astronomy,
the algorithm of HHT is briefly introduced in Section 2. Its
application to the DM time series of 30 pulsars to decompose the
general trends, annual and stochastic components is then presented in
Section 3. Discussions on the decomposed components and the
conclusions are given in Section 4 and 5, respectively.

\begin{figure*}
  \includegraphics[angle=0, width=0.78\textwidth] {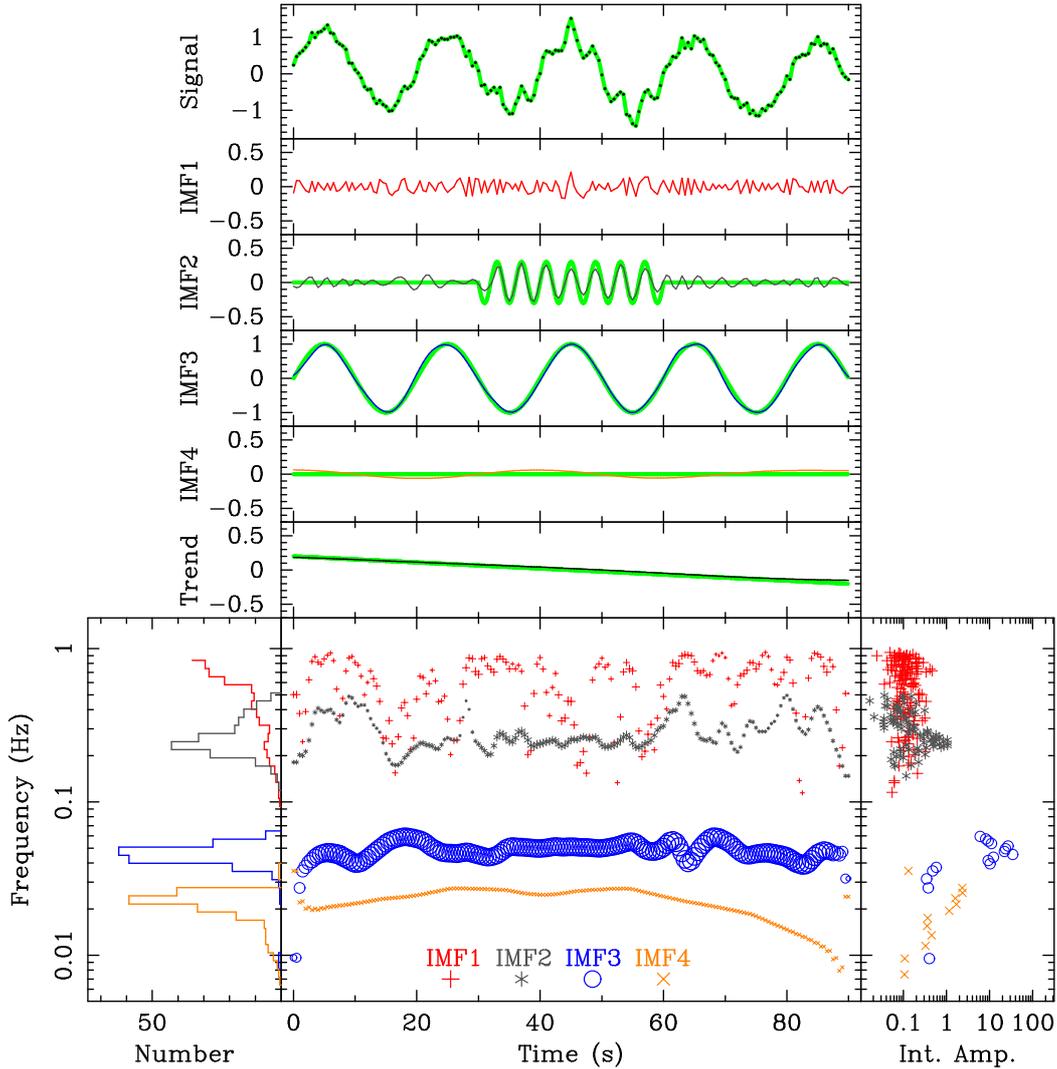}
  \caption{Demonstration of the HHT for a simulated time series. {\it
      The upper panels show the simulated data and IMFs from the
      EEMD}. The simulated time series are represented by the curve
    and 300 uniformly sampled data points in the top panel. The signal
    is originally comprised of four components: a linear term of
    $0.2-0.4~t/90.0$ (the same line in the trend panel), a low
    frequency oscillation of $1.0\sin(2\pi/20.0~t)$ (the curve in the
    panel for IMF3), a high frequency oscillation of
    $0.3\sin(2\pi/4.0~t)$ appearing between 30~s and 60~s (the curve
    in the panel for IMF2), and a normally distributed random noise
    with $\sigma=0.1$. The sampled signal is decomposed by the HHT
    into four IMFs and the trend, i.e., $\rm
    Signal=\sum_{i=1}^4{IMFi}+Trend$: here IMF1 is the noise, IMF2 the
    high frequency oscillation, IMF3 the low frequency oscillation,
    and IMF4 the residual of a small amplitude. {\it The bottom panels
      are the instantaneous frequencies and amplitudes from the
      Hilbert transform of the four decomposed IMFs}, i.e., each point
    in a given IMF has a corresponding pair of amplitude and
    frequency. The bottom middle panel is an amplitude-frequency-time
    plot. The time and frequency are represented by the horizontal and
    vertical axes. Each IMF has a dominating frequency interval, as
    shown by the histogram of instantaneous frequencies in the bottom
    left panel with the peak indicating the most probable
    frequency. By integrating the instantaneous amplitudes over time,
    we obtain the distribution of integrated amplitudes over frequency
    for each IMF, i.e. the marginal spectrum, as shown in the bottom
    right panel.}
  \label{fig:simu}
\end{figure*}

\input tab1.tex

\section{The Hilbert Huang Transform}

The HHT has been developed to process nonlinear and non-stationary
signals \citep{hsl+98,hsl99}. It has been used in many research areas
already, e.g. in the area of geophysics and meteorology for
decomposing the ionospheric scintillation effects from the global
navigation satellite system signals \citep[e.g.][]{ssv17} or
retrieving wind direction from rain-contaminated X-band nautical radar
sea surface images \citep[e.g.][]{lhg17}, in the area of Solar physics
for analyzing the Sunspot Numbers \citep[e.g.][]{gao16}, and in the
area of astrophysics for analyzing gravitational waves from the late
inspiral, merger, and post-merger phases of binary neutron stars
coalescence \citep[e.g.][]{kot+16}. However, it has not heretofore
been used for any analysis of real pulsar data.

The HHT consists of ``empirical mode decomposition'' (EMD) and the
well-known Hilbert transform. The EMD can decompose any complicated
data set into a finite and often small number of ``intrinsic mode
functions'' (IMFs) without a priori basis unlike Fourier-based
methods. These IMFs are generally in agreement with physical signal
interpretations, and hence the method can give sharp identifications
of embedded structures. These IMFs are then transformed to the Hilbert
spectrum to demonstrate the energy-frequency-time distribution of the
signal.

To calculate the EMD of a given signal, $x(t)$, its local maxima and
minima are firstly identified, and then the envelops for both types of
extremes are constructed. A mean curve is calculated by averaging both
envelops, which is then subtracted from the signal. This procedure is
called ``sifting'', and iteratively done many times until the
remaining signal meets the following criteria: 1) the number of
extremes and the number of zero crossings are equal or differ by one;
2) the mean for the envelops is zero. After this iterative process,
the finest component of the signal, i.e. the first intrinsic mode
function (IMF1), $c_1(t)$, is decomposed, which shows very fast
variations depending on the sampling cadence.

After this, the first residual, $r_1(t)=x(t)-c_1(t)$, is computed,
which now serves as the input signal for re-doing the iterative
sifting processes to get IMF2, $c_2(t)$. Then, the second residual,
$r_2(t)=r_1(t)-c_2(t)$, is obtained as the input signal for $c_3(t)$,
etc. Such a decomposition procedure is iteratively done to get IMF1 to
IMF$n$, until the residual, $r_n(t)$, becomes monotonic or has only
one extremum, which is called the ``trend'' term. The original signal
thus can be expressed as,
\begin{equation}
  x(t)=\sum_{j=1}^n{c_j(t)}+r_n(t).
  \label{eq:dm}
\end{equation}
See the upper panels of Figure~\ref{fig:simu} for illustration of the
EMD of a simulated signal.

The EMD has been proved to be very useful in geophysics, solar physics
and other scientific fields, as mentioned above. However, it still has
some drawbacks. The most serious problem is the mode mixing, i.e. a
signal of similar scales and frequencies appears in different
IMFs. The Ensemble Empirical Mode Decomposition (EEMD), a noise
assisted data analysis method, was later developed by \citet{wh09} to
solve the problem, in which the independent white noise realizations
are performed and added to the original data to get an ensemble of
EMDs. The IMFs of an ensemble of EMDs are then averaged to eliminate
the added white noise, by which the mixed modes can be separated.

For each of the decomposed IMFs, the Hilbert transform can be applied
to obtain their instantaneous frequencies and amplitudes. The Hilbert
transformation of a signal, $x(t)$, can be written as:
\begin{equation}
  y(t)=\frac{P}{\pi}\int_{-\infty}^{-\infty}{\frac{x(\tau)}{t-\tau} d\tau}.
  \label{eq:hil}
\end{equation}
Here, $P$ is the Cauchy principal value of the signal integration. The
complex signal then reads
\begin{equation}
  z(t)=x(t)+iy(t)=a(t)\exp^{i \theta(t)},
  \label{eq:complex}
\end{equation}
and $a(t)$ and $\theta(t)$ represent the instantaneous amplitude and
phase. The instantaneous frequency is defined as
\begin{equation}
  \omega(t)=\frac{d \theta}{d t}.
  \label{eq:freq}
\end{equation}
Instantaneous frequencies and amplitudes of the IMFs demonstrate the
energy-time-frequency distribution for the input signal $x(t)$, as
shown in the bottom panels of Figure~\ref{fig:simu}. The EMD or EEMD
and Hilbert spectral analysis are combined together to form the
Hilbert-Huang Transform (HHT). The open tools of the HHT are available
at the
web-page\footnote{https://cran.r-project.org/web/packages/hht/index.html}.

\section{HHT analysis of pulsar DM variations}

We apply the HHT method to the temporal variations of pulsar DMs.  DM
data of 30 pulsars, called ``DMX'' standing for the offsets from the
formal DM values in the ephemerides, are taken from
\citet{abb+18}. These pulsars have been observed for the North
American Nanohertz Observatory for Gravitational Waves project, and
more than 48 DM measurements over more than 3.8 years are available
(see Table~\ref{table:hht}) for each one. We firstly decompose the DM
time series directly by EEMD for the long term ``{\it general trend}''
(not just the trend from the HHT, see below) and the short term
``random noise'', and then obtain the annual DM variations by folding
the trend- and noise-subtracted data.

\begin{figure*}
  \centering
  \includegraphics[bb = 55 33 328 365, clip, height=0.463\textheight] {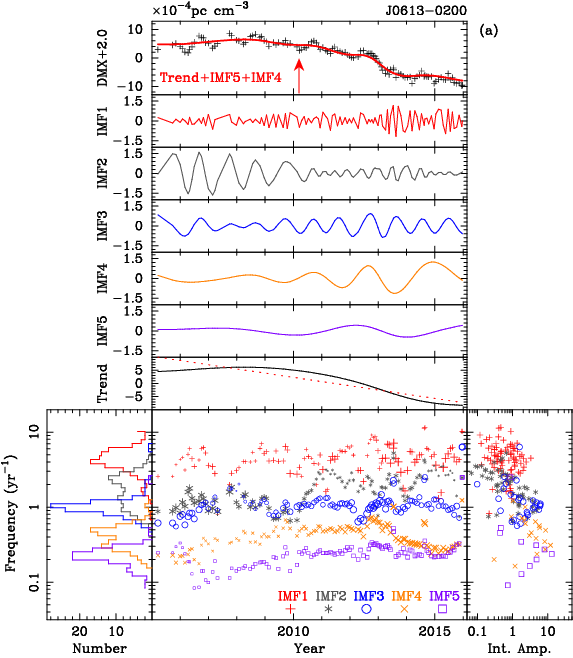}
  \includegraphics[bb = 75 33 328 365, clip, height=0.463\textheight] {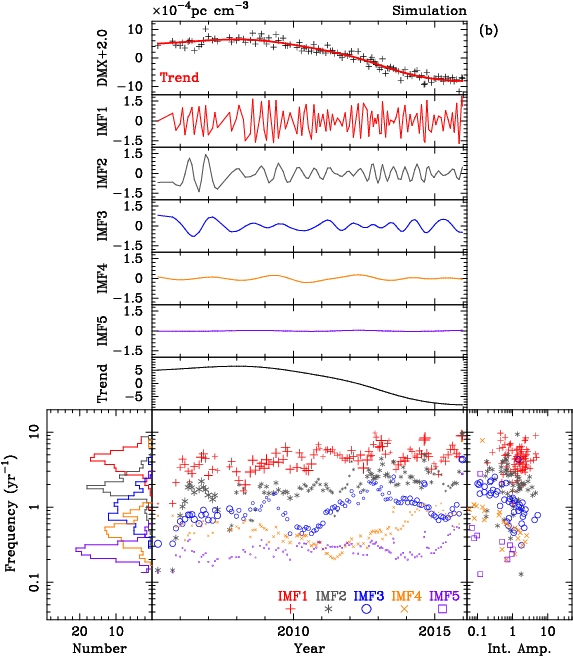}
  \caption{The HHT for DM variations of PSR J0613-0200. (a) The EEMD
    and Hilbert spectral analysis of observed DM variations are
    demonstrated in the upper and bottom panels, respectively. {\it In
      the upper panels,} the temporal DM variations (i.e. DMX, in
    units of $10^{-4}$~pc~cm$^{-3}$) are represented by ``+'' in the
    top panel. The measurements after the epoch indicated by red arrow
    have a higher observation cadence and used for folding for the
    annual variation term. The EEMD results, i.e., IMF1 to IMF5 and
    the trend, are demonstrated in red, grey, blue, orange, purple and
    black in the panels downward. The trend is fitted by a dotted line
    to get an average gradient of DM variations as listed in
    Table~\ref{table:hht}. The {\it general trend}, not just the EEMD
    trend term, are composited by "Trend+IMF5+IMF4", representing
    variations with periodicity longer than one year. Instantaneous
    frequencies and amplitudes of the IMFs for the Hilbert spectrum
    are shown {\it in the bottom panels}. The amplitude-frequency-time
    plot, histogram for the instantaneous frequencies and the marginal
    spectrum for each IMF are shown in the {\it middle, left and right
      panel}, respectively, see the keys in Fig.~\ref{fig:simu}. (b)
    The EEMD and Hilbert spectra of the simulated DM variations, which
    have the same cadence as real observations of this pulsar and do
    not show an annual variation term from EEMD.}
  \label{fig:J0613DMX}
\end{figure*}

\begin{figure*}
  \centering
  \includegraphics[angle=0, width=0.77\textwidth] {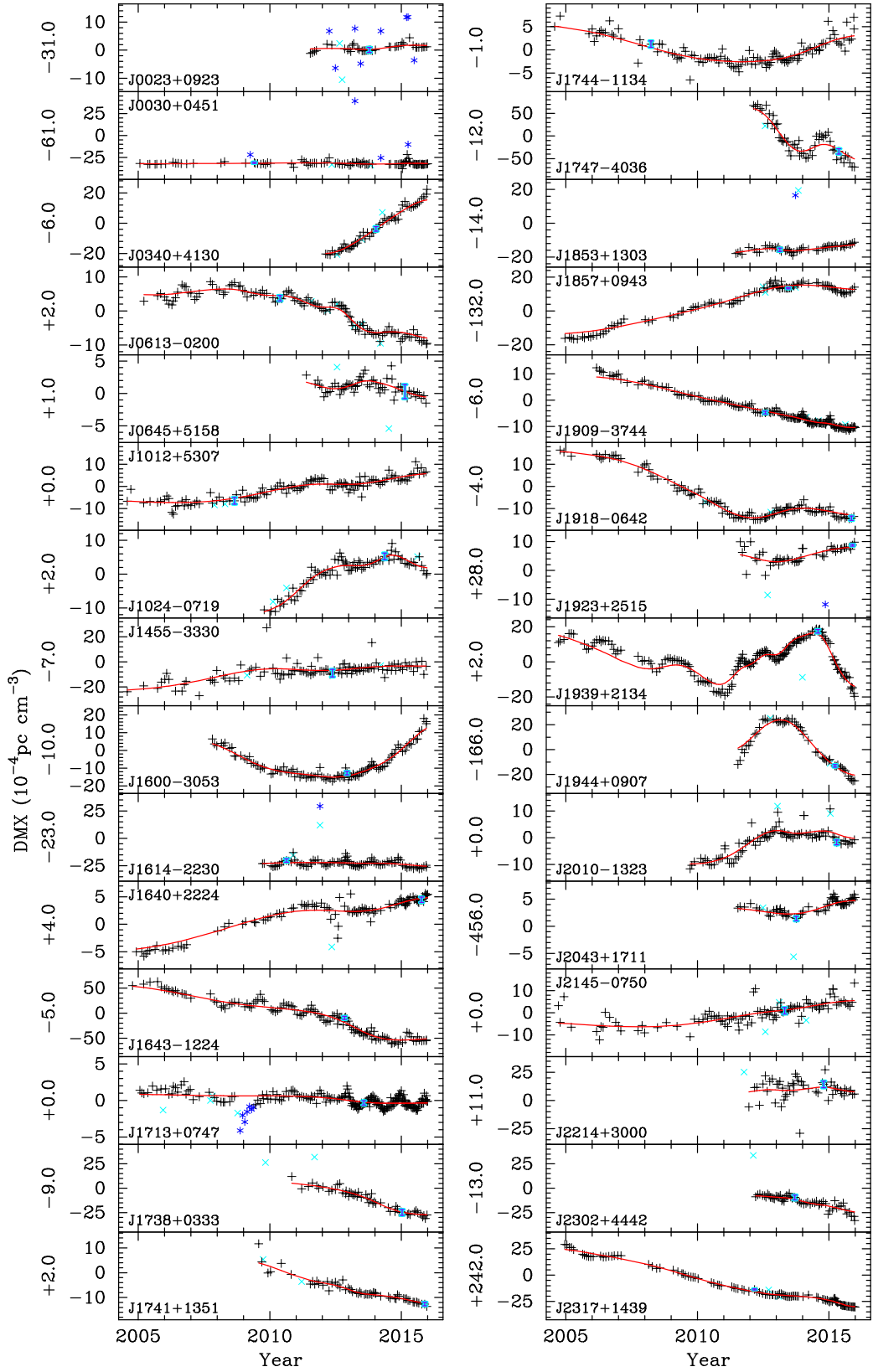}
  \caption{Temporal DM variations of 30 pulsars and the {\it general
      trends} represented by the curves. Normal data are indicated by
    plus, DM measurements with uncertainties three times larger than
    their median are indicated by "$\times$'' and discarded in the
    analysis. The data deviating from their neighbors by more than ten
    times the median variation, as indicated by the asteroids, are
    also omitted in the HHT analysis, for example, the steep drop and
    recovery of DM variation for PSR J1713+0747 caused by scattering.}
  \label{fig:trend}
\end{figure*}

\subsection{EEMD for the general trend and noise}

DMXs are very small deviations from the formal DM values of pulsars
and scaled to the units of $10^{-4}$~pc~cm$^{-3}$. Through EEMD, DMX
time series can be decomposed: DMX = $\sum_{i=1}^N$ {\rm IMFi} +
trend. Figure~\ref{fig:J0613DMX} shows an example of the EEMD for
IMFs.

It is noticed that each DMX measurement has an uncertainty. The
uncertainties for data obtained in early years are much larger than
those in the recent years because of lower sensitivity of the old
observing systems with limited observing bandwidth. Since the
uncertainties are not concerned in EEMD, we have to ``clean'' the DMX
data before the EEMD process. We first omit those data with a very
large uncertainty (three times larger than the median uncertainty),
which are often deviated from the general trend \citep[see][]{abb+15}.
Some data are occasionally very (ten times larger than the median
derivation) deviated from their adjacent measurements as caused by the
very sharp annual DM variations (e.g. J0023+0923) or the extreme
scattering event (e.g. PSR J1713+0747) \citep{abb+15, cks+15, lsc+16},
which are also not the proper tracers for the trend. If these data
points are otherwise included, the HHT analysis will introduce a
number of oscillations with various amplitudes and frequencies to the
IMFs at their instants.

HHT analysis of the ``cleaned'' DMX data gives IMF1 to IMF5 and the
trend, as shown for the DM time series for PSR J0613$-$0200 in
Figure~\ref{fig:J0613DMX}. The ``noise'' term is {\rm IMF1} as seen in
the upper panels of Figure~\ref{fig:J0613DMX}(a), which represents the
finest structure of DM variation with a time scale depending on the
observational cadence. The Hilbert transform of {\rm IMF1} gives its
instantaneous frequencies and amplitudes, as shown by red dots in the
bottom middle panel of Figure~\ref{fig:J0613DMX}(a). The median values
of instantaneous amplitudes and frequencies as well as their standard
deviations can be found for this random noise component of DM
variations, as listed in columns (9) and (10) in
Table~\ref{table:hht}. IMF3 and IMF2 are annual and semi-annual
variations, which will be analyzed in the next section.

The EEMD trend term for DM variations of PSR J0613$-$0200 is not
linear but can still be fitted approximately with a line to get the
rough rate of DM changes, i.e. ${\rm d} {\rm DM}/{\rm d}t$, as also
listed in column (8) of Table~\ref{table:hht}. The {\it general trend}
term for DM variations of PSR J0613$-$0200, however, is not just the
EEMD trend term, but is composited by "Trend+IMF5+IMF4" which shows
long-term variations with time-scales longer than one year, as
indicated the line in the top panel.

\begin{figure*}
  \centering
  \includegraphics[angle=0, width=0.73\textwidth] {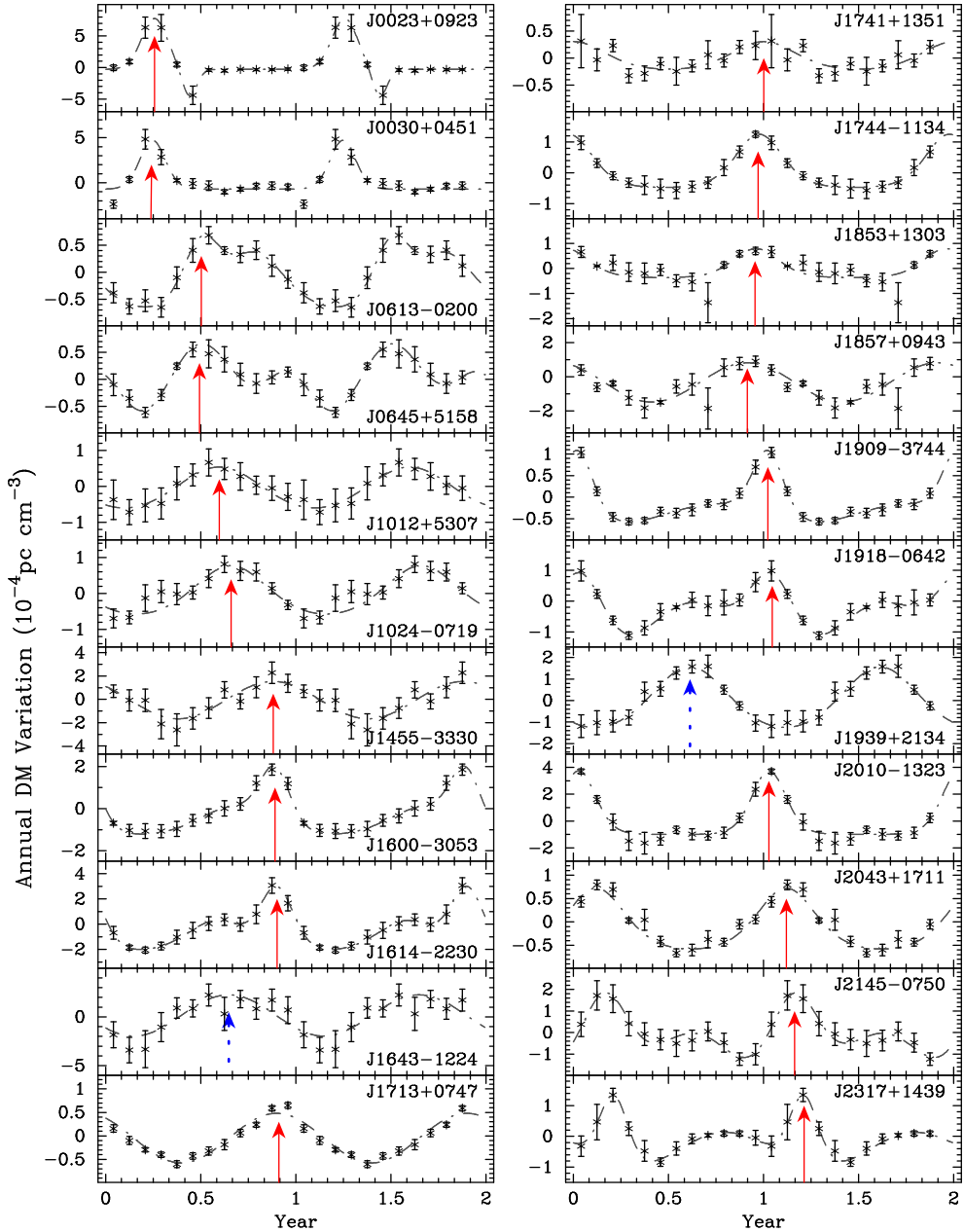}
  \caption{Annual DM variations of 22 pulsars, obtained by folding
    data obtained in recent years with one year period (referring to
    the beginning of a year) after the ``general trend'' and noise
    terms are subtracted (see plots for DM variations in
    Figure~\ref{figA}). Two cycles are plotted for clarity. The fitted
    von Mises (circular normal) curves are plotted by dash-dotted
    lines, and the peak phases of annual variations are indicated by
    arrows. Two pulsars, PSRs J1643$-$1224 and J1939+2134, have
    extraordinary peak phases as indicated by dashed arrows.}
  \label{fig:ann}
\end{figure*}

To demonstrate the effectiveness of the HHT in decomposing DM
variations, we simulate a DM time series by using the trend term of
Figure~\ref{fig:J0613DMX}(a) plus the white noises. The simulated data
in Figure~\ref{fig:J0613DMX}(b) have the same cadence as the real
observations for PSR J0613$-$0200. The HHT decomposes the simulated
data into five IMFs and trend term. Though each IMF has a dominating
frequency range, but is not sharply peaked at any given frequency
(e.g. one year$^{-1}$). The marginal spectra have comparable power
among these IMFs, rather than a significant power excess for a given
IMF. It is also noticed that the phases for the peaks of a given IMF
(e.g. IMF2 and IMF3) vary a lot, rather than being fixed at a given
phase over years. In other words, these IMFs of simulated data do not
show any significant periodic signal, which demonstrate that HHT does
not artificially introduce any regular annual signals in the DM time
series.

As shown in Appendix A, we have also decomposed the noise terms and
the {\it general trends} for 30 pulsars (see Figure~\ref{figA}). The
{\it general trends} have also been plotted together with data in
Figure~\ref{fig:trend}.

\subsection{Annual variations from folding DMX data}

There is no doubt that annual variations exist for pulsar DMs, which
have been shown by clear frequency peaks for IMF2 and IMF3 of PSR
J0613$-$0200 in Figure~\ref{fig:J0613DMX} (a). Observations with
higher cadence since the year of about 2010 (see Figure~\ref{figA})
exhibit clear semi-annual and annual variations.

Annual variations do not have to follow the sinusoidal curve
\citep{lcc+16}. They can have frequencies of annual and semi-annual
and other forms as shown by Figure~\ref{fig:J0613DMX} (a). Such an
annual variation can be obtained by adding IMF2 and IMF3. The most
effective approach to get the annual term is folding the DMX curves
with an one-year period, after the general trend and noise terms are
subtracted from the original data. Data of the recent observations
after the epoches indicated by the arrows in
Figure~\ref{fig:J0613DMX}(a) and Figure~\ref{figA} are folded into 12
bins, corresponding to 12 months a year. DMX data in each bin are
weighting averaged according to the measurement uncertainty. Among the
30 pulsars, 7 pulsars show complicated structures in the trend- and
noise-subtracted data, and no annual variations can be identified
within short data spans. The other one, PSR J1640+2224, shows a
complicated feature (see Figure~\ref{figA}).  The resulting annual
curves for the remaining 22 pulsars are displayed in Figure~\ref{figA}
and Figure~\ref{fig:ann}.

As shown in Figure~\ref{fig:trend}, data deviating from the adjacent
measurements by more than ten times the median derivation were omitted
for the trend analysis for 5 pulsars. Most of these ``cleaned'' data
are around the peaks for annual variations (see Figure~\ref{figA} for
PSRs J0023+0923, J0030+0451 and J1614$-$2230). Outlying data indicated
by asteroids near the curve peak are taken back to form the annual
variation curves, if they do not significantly differ from data in
other peaks. Though near each peak only one measurement is available,
the repeatedly outlying data near the peak of annual variations for
PSRs J0023+0923 and J0030+0451 indicate that the real peak should be
much more sharp than we see from merely one measurement each. For PSRs
J1614$-$2230, J1853+1303 and J1923+2515, only one data point is
abnormal near the peak, which is not taken back because no recurrence
has been observed for confirmation. The data points for extreme
scattering event of PSR J1713+0747 are certainly not included to fold
for the annual curve.

To quantitatively describe the annual variations, we fit the von Mises
functions (i.e. the circular normal function) to the folded annual
curves. For the curves for PSRs J0023+0923, J0613$-$0200, J0645+5158,
J1600$-$3053, J1614$-$2230, J1909$-$3744, J1918$-$0642, J2145$-$0750,
and J2317+1439, a single von Mises function is not enough and two von
Mises functions are employed for the fitting. The peak phases are
referenced to the beginning of a year, and can be so-obtained during
the fitting as indicated by the arrows in Figure~\ref{fig:ann}. The
amplitudes of annual variations are simply taken as the difference
between the maximum and minimum of the folded data. These values are
listed in columns (11) and (10) in Table~\ref{table:hht},
respectively.

\begin{figure*}
  \centering
  \includegraphics[angle=0, width=0.70\textwidth] {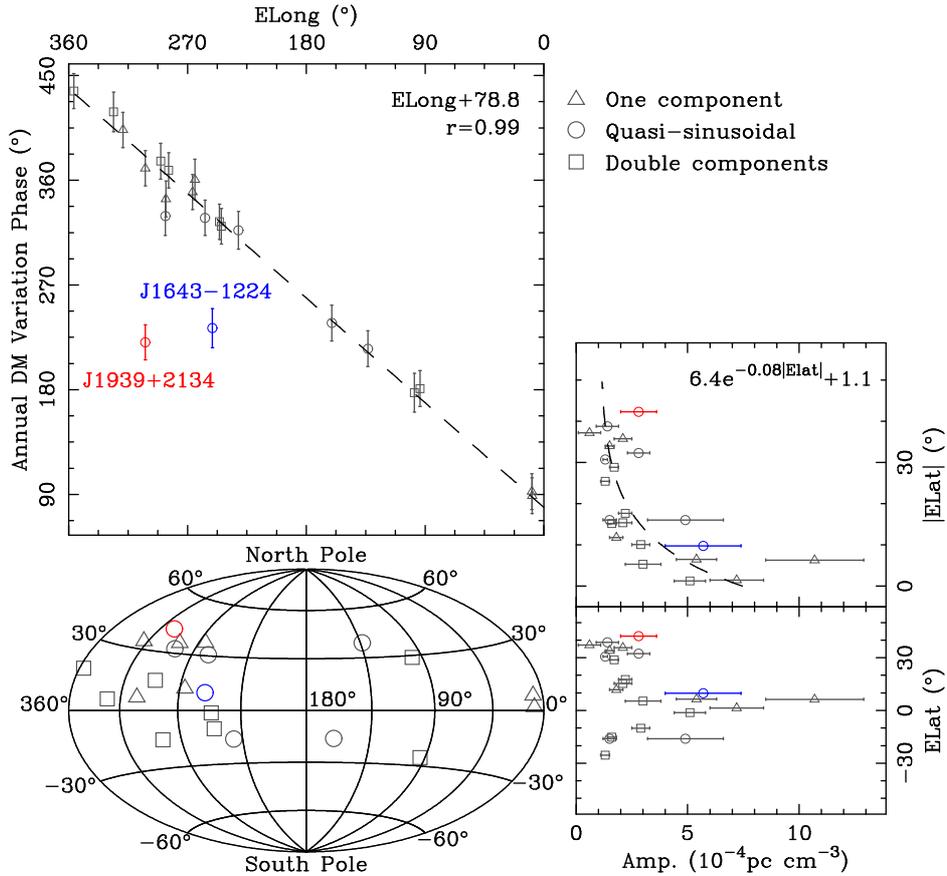}
  \caption{Distribution of 22 pulsars with respect to the ecliptic
    frame ({\it bottom left panel}). The phases and amplitudes of
    annual variations are related to the ecliptic longitude and
    latitude in the {\it upper} and {\it right} panels. Three kinds of
    annual variations, one component, quasi-sinusoidal and double
    components features, are represented by triangle, circle and
    rectangle points, respectively. {\it Right panels:} Amplitudes of
    annual variations are related to the ecliptic latitude or their
    absolute values ({\it upper panel}).  {\it Top panel:} Phases of
    annual variations are well correlated to the ecliptic longitude,
    with a correlation factor of $r=0.99$. The best fittings for the
    dependencies are indicated by the dashed lines. Two exceptions
    (PSRs J1643$-$1224 and J1939+2134) are marked.}
  \label{fig:ann_dis}
\end{figure*}

\section{Discussions}

We have decomposed the temporal DM variations of 30 pulsars into
general trends and small-amplitude random fast DM changes by using the
EEMD of the HHT, and then obtained the annual variation curves of 22
pulsars by folding the trend- and noise-subtracted DMX data.

\subsection{The trend and noise}

The general trends exhibit monotonic increasing, decreasing, or
complicated variations (see Figure~\ref{fig:trend} for the 30
pulsars). Clear monotonic DM decreasement has been observed for PSRs
J1614$-$2230, J1643$-$1224, J1713+0747, J1738+0333, J1741+1351,
1909$-$3744, J2302+4442 and J2317+1439, clear monotonic DM
increasement for PSRs J0340+4130, J1012+5307 and J1455$-$3330, and
quadratic variations for PSRs J1600$-$3053 and J1944+0907. The ${\rm d
  DM}/{\rm d}t$ values in Table~\ref{table:hht} obtained by the simple
linear fitting to the EEMD trend term can only roughly reflect the
averaged DM gradient, and listed there for comparison with those in
\citet{jml+17} from the 9 year data-set of \citet{abb+15}. PSRs
J0340+4130, J1643$-$1224, J1747$-$4036 and J1944+0907 have the largest
gradients for DM variations, more than $8 \times 10^{-4} {\rm cm}^{-3}
{\rm pc}$ per year. As discussed in \citet{lcc+16}, the relative
line-of-sight motion between the Earth and pulsar can produce
monotonically varying DM if density gradients of interstellar medium
are not taken into account. Otherwise, plasma wedges can also cause
linear trends \citep{bhv+93}. In most cases, the DM variations exhibit
much complicated trends. For example, PSR J1939+2134 has a DM
decreased before 2011 and increased afterwards to 2014, after which it
decreases again. PSR J1747$-$4036 demonstrates similar variation
features, but with much shorter time scale. They may be caused by the
transverse motion of a large interstellar cloud into or out of the
line of our sight. Moreover, ionized bow shocks and stochastic process
of the interstellar medium with a red power spectrum may also cause
non-monotonic trends for the DM variations \citep{lcc+16}.

By performing EEMD on the DM time series, we got the random terms of
DM variations, as shown by IMF1 in Figure~\ref{fig:J0613DMX} (a) for
PSR J0613$-$0200 and the middle panel in each plot of
Figure~\ref{figA} for other pulsars. The medians for the instantaneous
frequencies of IMF1s range from 3.6 to 10.6 per year as listed in
Table~\ref{table:hht}, which in fact reflect only the main cadence of
observations. This ``noise'' part of the observed DMs comes from
random contributions from interstellar medium, interplanetary medium
and ionosphere, but it should have turbulence at different physical
and hence time scales \citep{ars95}. The small-scale turbulence in
interstellar medium can cause much faster DM variations, such as the
extreme scattering-like variation of PSR J1713+0747 around 2009. For
the Kolmogorov turbulence, it was predicted that the power spectrum of
the DM variations has a spectral index of 8/3 \citep{lcc+16}.  More
frequent observations in future can reveal better the turbulence at
different scales in the interstellar medium, interplanetary medium and
ionosphere. In addition, EEMD also demonstrates low frequency
variations with periodicity longer than one year, which may be mixed
with the trend term. Though HHT can decompose DM variations into a
number of components with different time scales, it is very premature
to use them to interpret the detailed structures of interstellar
medium.

\subsection{Annual DM variations}

Annual pulsar DM variation caused by the variable interplanetary
medium and Solar wind has been known for a long time
\citep[e.g.][]{yhc+07}. \citet{kcs+13} found an annual periodicity
for PSRs J0613$-$0200, J1045$-$4509, J1643$-$1224 and J1939$+$2134
from spectral analysis of DM variations obtained by the Parkes Pulsar
Timing Array project. \citet{lcc+16} obtained the annual variation of
PSR J1909$-$3744 by subtracting a linear trend from the DM
measurements. \citet{jml+17} detected annual DM variations for an
ensemble of pulsars from the 9-year data set of \citet{abb+15} by
decomposing the time series with annual triangle functions and
Lomb-Scargle periodogram analysis.

We obtained the annual variation curves for 22 pulsars (see
Figure~\ref{fig:ann} and Figure~\ref{figA}) especially by using the
data of high quality recent observations from \citet{abb+18}. Compared
with previous analyses, annual variation curves of 9 pulsars (PSRs
J0023+0923, J1012+5307, J1024-0719, J1455$-$3330, J1600-3053,
J1713+0747, J1741+1351, J1853+1303, and J2145-0750) are obtained here
for the first time, and the curves for other pulsars are significantly
improved.

As seen from Figure~\ref{fig:ann}, annual variations exhibit three
kinds of features. A simple quasi-sinusoidal variation has been
obtained for PSRs J1012+5307, J1024$-$0719, J1455$-$3330,
J1643$-$1224, J1741+1351, J1857+0943 and J1939+2134. These pulsars are
generally away from the ecliptic plane (see Figure~\ref{fig:ann_dis}),
and the quasi-sinusoidal variations can be understood by the
trajectory of Earth in the interplanetary medium \citep{kcs+13}.
One peak component of variations can be seen for PSRs J0023+0923,
J0030+0451, J1713+0747, J1744$-$1134, J1853+1303, J2010$-$1323 and
J2043+1711. These pulsars are located at low ecliptic latitudes and
seriously affected by electron density contributed by the solar wind.
The strange DM ``absorption'' after the peak for PSR J0023+0923 has
been detected in three years (see Figure~\ref{figA}), and therefore it
is confirmed but not explainable. The ``absorption'' before the peak
of PSR J0030+0451 is still to be confirmed.
Double peaks of variations can be seen for PSRs J0613$-$0200,
J0645+5158, J1600$-$3053, J1614$-$2230, J1909$-$3744, J1918$-$0642,
J2145$-$0750, and J2317$+$1439, which may result from the coupled
effects of the ionosphere together with the interplanetary medium.

With the annual variations of a large sample of 22 pulsars, we can
check the correlations of both the amplitudes and phases of annual
variations with the ecliptic coordinates, as shown in
Figure~\ref{fig:ann_dis}.  In general the higher-density slow wind
appears at lower ecliptic latitude and the lower-density fast wind at
higher latitude. Because the electron density generally scales with
distance $r$ from the Sun by following $r^{-2}$, the amplitudes of
annual variations should also vary with ecliptic latitude. The
correlation of annual variations with solar position angle has been
analyzed for some individual pulsar previously, for example a strong
correlation was found for PSRs J0030+0451 and J1614-2230 and a
moderate correlation for PSR J2010-1323 \citep{jml+17}. The DM
variations of PSR J1909$-$3744 have been analyzed by \citet{lcc+16}
with modeling the annual variations by the solar wind as well as
ionosphere and heliosphere. We found from our new curves that pulsars
closer to the ecliptic plane tend to have larger amplitudes for annual
variations, which can be roughly described by an exponential
function. The peak phases of annual variations are well correlated
with the ecliptic longitude, with a correlation coefficient $r=0.99$,
if two exceptions (PSRs J1643$-$1224 and J1939+2134) are not
considered.

Apparent annual variations for PSRs J1643$-$1224 and J1939+2134 are
very different. Their amplitudes are somehow larger than the expected
values from the fitted exponential function curve. Their peak phases
are very exceptional, which indicate that annual variations of both
pulsars might not only result from the solar wind. It should be noted
that the amplitude for annual variation of PSR J1643$-$1224 is very
large (up to $5.7 \times 10^{-4} {\rm cm}^{-3} {\rm pc~}$) and the
trend curve for PSR J1939+2134 is the most complicated. These
``annual'' variations may be caused by quasi-periodically crossing
of the interstellar cloud by sight lines, so that their phases can be
arbitrary \citep{lcc+16}. By jointly investigating temporal variations
in DM and scattering in the ionized interstellar medium through the
Bayesian approach, \citet{lkd+17} also suggested that the scattering
variation of interstellar clouds is the most possible cause for the
period apparent DM variation observed in PSR J1643$-$1224.

\section{Conclusions}

We decomposed the temporal DM variations of 30 pulsars into general
trends, random components and annual variations by employing the HHT
method. We found that HHT can extract the general trends from data,
which represent the largest structures of DM variations resulting from
the irregularly distributed interstellar medium. The random DM
components heavily depend on the cadence of observations. Annual
variations of 22 pulsars can be extracted from the trend- and
noise-subtracted DM data, which exhibit quasi-sinusoidal, one
component and double components features. The amplitudes and phases of
annual variations are well related to ecliptic latitude and longitude,
respectively, whose origins are attributed to the DM variations from
solar wind and interplanetary medium. The exceptions are probably
caused by the interstellar medium.

In future, observations with higher cadence can help us to understand
the turbulent feature of the interstellar medium. Long-term
observations can be used to reveal the long-term variations due to the
interstellar refractive scattering or the annual variations caused by
the solar wind and interplanetary medium. Therefore, sensitive
wide-band observations of more pulsars with higher cadence in longer
time spans are very important, which can not only show the high
precision DM variations in great details, but also reveal the details
of the interplanetary and interstellar medium. In addition, the HHT
can also be implemented into the Bayesian pulsar timing algorithms to
check if the timing noises have variations on different temporal
scales.

\section*{Acknowledgements}

The authors thank Profs. Kinwah Wu, W.~A. Coles, W.~W. Zhu and also
the anonymous referee, for their helpful discussions and
comments. This work has been partially supported by the National
Natural Science Foundation of China (11403043, 11473034), the Young
Researcher Grant of National Astronomical Observatories Chinese
Academy of Sciences, the Key Research Program of the Chinese Academy
of Sciences (Grant No. QYZDJ-SSW-SLH021), the strategic Priority
Research Program of Chinese Academy of Sciences (Grant
No. XDB23010200) and the Open Project Program of the Key Laboratory of
FAST, NAOC, Chinese Academy of Sciences.

\bibliographystyle{mnras}
\bibliography{hht}

%

\appendix

\section{Plots for the trend, random, and annual variations for pulsar DMs}

We present the temporal DM series and the general trends, random fast
variations and annual variation components for all 30 pulsars.

\begin{figure*}
  \label{figA}
  \includegraphics[angle=0,height=0.15\textheight] {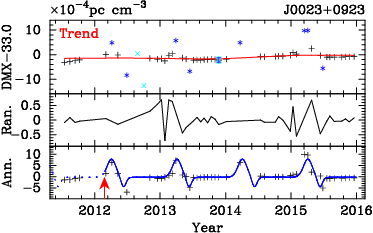}
  \includegraphics[angle=0,height=0.15\textheight] {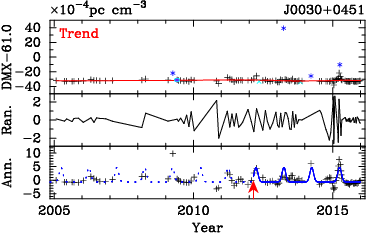}
  \includegraphics[angle=0,height=0.15\textheight] {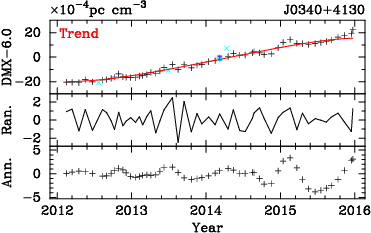}\\[1mm]
  \includegraphics[angle=0,height=0.15\textheight] {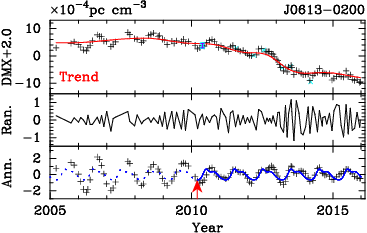}
  \includegraphics[angle=0,height=0.15\textheight] {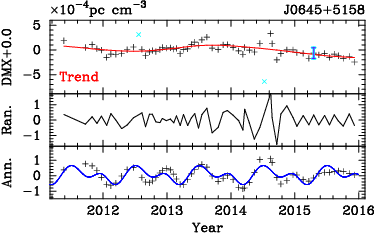}
  \includegraphics[angle=0,height=0.15\textheight] {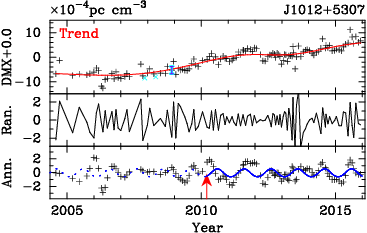}\\[1mm]
  \includegraphics[angle=0,height=0.15\textheight] {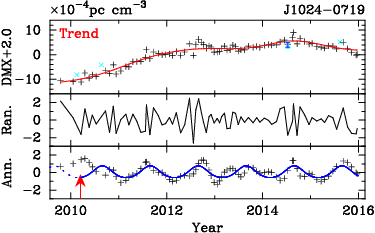}
  \includegraphics[angle=0,height=0.15\textheight] {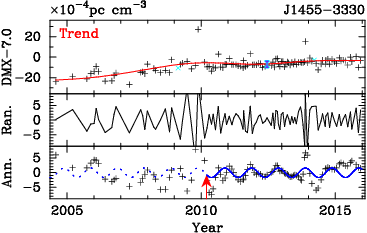}
  \includegraphics[angle=0,height=0.15\textheight] {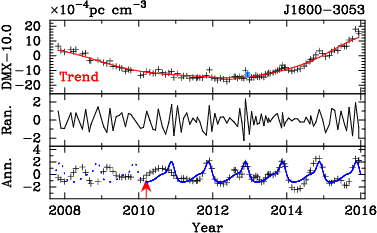}\\[1mm]
  \includegraphics[angle=0,height=0.15\textheight] {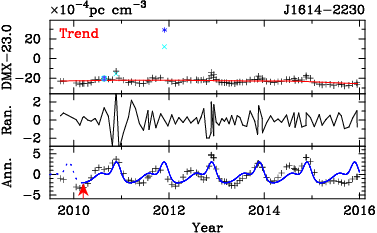}
  \includegraphics[angle=0,height=0.15\textheight] {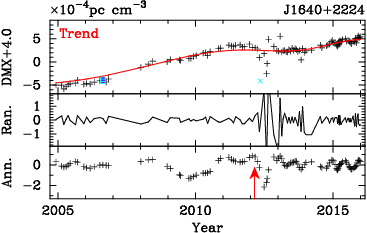}
  \includegraphics[angle=0,height=0.15\textheight] {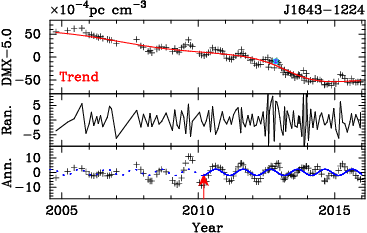} \\[1mm]
  \includegraphics[angle=0,height=0.15\textheight] {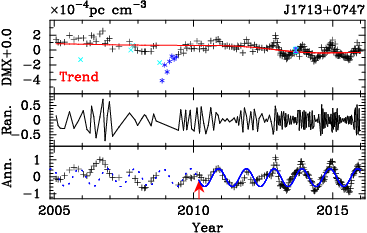}
  \includegraphics[angle=0,height=0.15\textheight] {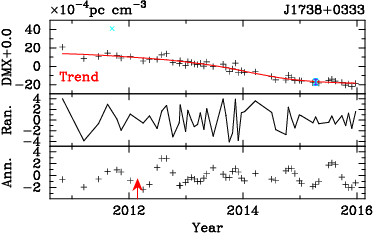}
  \includegraphics[angle=0,height=0.15\textheight] {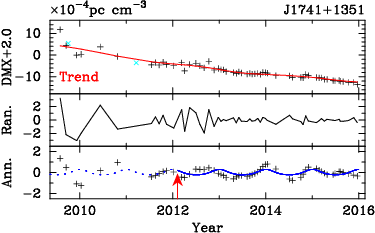} \\[1mm]
  \caption{For each pulsar, the original DMX data are shown in the top
    panel together with the ``general trend''. The data are plotted in
    different symbols as in Figure~\ref{fig:trend}: normal data by
    ``+''; measurements with uncertainties three times larger than the
    median uncertainty are indicated by ``$\times$''; the data
    deviating from their neighbors by more than ten times the median
    deviation are indicated by the asteroids. The middle panel is
    plotted for the random fluctuations obtained as IMF1 by the EEMD,
    with simply connected lines. The data with ``general trend'' and
    noise subtracted are then plotted in the bottom panel, with the
    annual variation curves plotted in the solid line after the epoch
    indicated by the arrow after which data were used for folding for
    the annual curves. The curves are extended to early data by the
    dotted line. Outlying data indicated by asteroids near the curve
    peaks are taken back to form the annual variation curves, if they
    do not differ significantly from the data at other peaks.  }
\end{figure*}

\begin{figure*}
  \centering
  \addtocounter{figure}{-1}
  \includegraphics[angle=0,height=0.15\textheight] {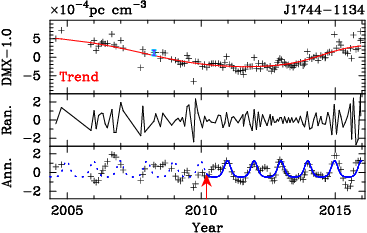}
  \includegraphics[angle=0,height=0.15\textheight] {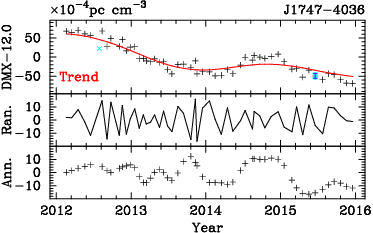}
  \includegraphics[angle=0,height=0.15\textheight] {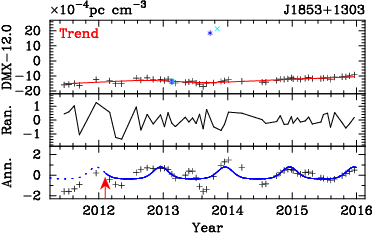} \\[1mm]
  \includegraphics[angle=0,height=0.15\textheight] {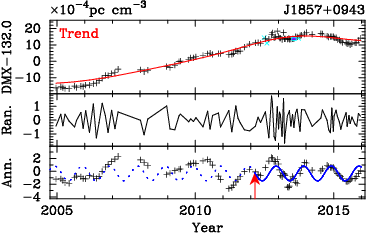}
  \includegraphics[angle=0,height=0.15\textheight] {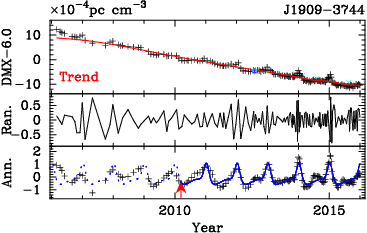}
  \includegraphics[angle=0,height=0.15\textheight] {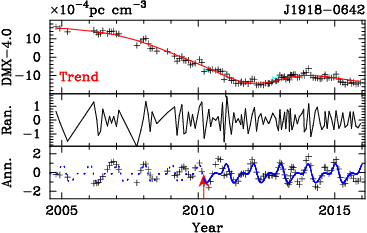} \\[1mm]
  \includegraphics[angle=0,height=0.15\textheight] {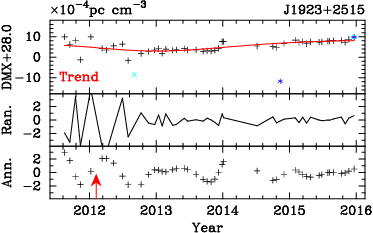}
  \includegraphics[angle=0,height=0.15\textheight] {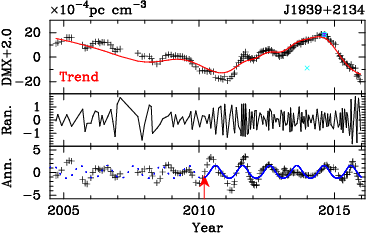}
  \includegraphics[angle=0,height=0.15\textheight] {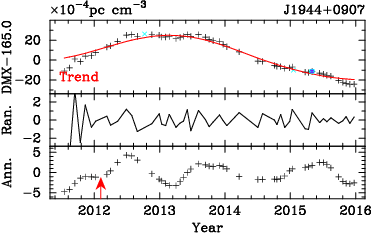} \\[1mm]
  \includegraphics[angle=0,height=0.15\textheight] {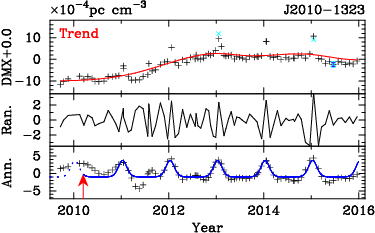}
  \includegraphics[angle=0,height=0.15\textheight] {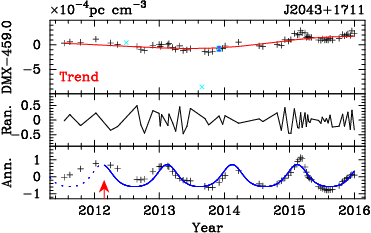}
  \includegraphics[angle=0,height=0.15\textheight] {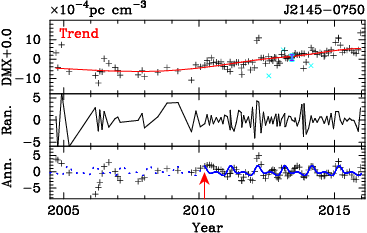}\\[1mm]
  \includegraphics[angle=0,height=0.15\textheight] {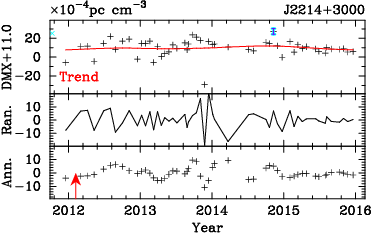}
  \includegraphics[angle=0,height=0.15\textheight] {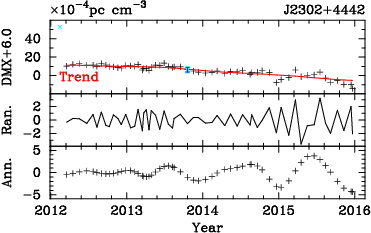}
  \includegraphics[angle=0,height=0.15\textheight] {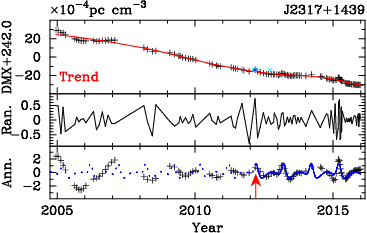}

  \caption{{\bf Continued--}}
\end{figure*}

\label{lastpage}
\end{document}

%% file: tab1.tex
\begin{table*}
  \caption{Observational parameters for DM variations of 30 pulsars and derived parameters for the trend,
    noise and annual terms.}  \centering
  \label{table:hht}
  \small
  \begin{tabular}{rlrrrrc|r|cc|rr}
    \hline
    \hline
  No. & PSR name & E-Long. & E-Lat. & Obs. No. & Span & $\sigma_{\rm DMX}$ & \multicolumn{1}{c}{${\rm d DM}/{\rm d}t$} &
      IA$_{\rm noise}$ & IF$_{\rm noise}$ & $\rm \Delta DM_{\rm ann.}$ & Phase$_{\rm ann.}$  \\
 (1)&  \multicolumn{1}{c}{(2)} & \multicolumn{1}{c}{(3)}    &  \multicolumn{1}{c}{(4)}   &   (5)    &  (6) &    (7)  & \multicolumn{1}{c}{(8)} &
        (9)          &   (10)        &    \multicolumn{1}{c}{(11)}              & \multicolumn{1}{c}{(12)}           \\
      \hline
 1 & J0023$+$0923 &   9.07 &$   6.31$&  50 &  4.4 & 0.9 &$  0.37\pm$0.03 & 0.4$\pm$0.3 & 4.8$\pm$2.2 &10.7$\pm$2.2 &  92.4$\pm$15.3 \\ 
 2 & J0030$+$0451 &   8.91 &$   1.44$& 102 & 10.9 & 0.9 &$ -0.06\pm$0.01 & 1.0$\pm$0.8 & 5.9$\pm$4.3 &  7.2$\pm$1.2 &  88.9$\pm$15.3 \\ 
 3 & J0340$+$4130 &  62.61 &$  21.33$&  56 &  3.8 & 1.5 &$ 10.45\pm$0.16 & 1.3$\pm$0.6 & 5.1$\pm$2.0 &      -      &        -       \\ 
 4 & J0613$-$0200 &  93.80 &$ -25.41$& 121 & 10.8 & 0.8 &$ -1.61\pm$0.07 & 0.6$\pm$0.3 & 4.8$\pm$2.0 & 1.3$\pm$0.2 & 181.0$\pm$15.3 \\
 5 & J0645$+$5158 &  98.06 &$  28.85$&  61 &  4.5 & 1.1 &$ -0.29\pm$0.05 & 0.6$\pm$0.4 & 5.3$\pm$2.3 & 1.7$\pm$0.2 & 177.5$\pm$16.7 \\ 
 6 & J1012$+$5307 & 133.36 &$  38.76$& 123 & 11.4 & 1.5 &$  1.24\pm$0.01 & 1.4$\pm$0.9 & 4.6$\pm$2.5 & 1.4$\pm$0.5 & 215.2$\pm$15.4 \\
 7 & J1024$-$0719 & 160.73 &$ -16.04$&  82 &  6.2 & 1.0 &$  2.19\pm$0.13 & 1.3$\pm$0.8 & 4.8$\pm$2.1 & 1.5$\pm$0.3 & 237.3$\pm$15.4 \\
 8 & J1455$-$3330 & 231.35 &$ -16.04$& 108 & 11.4 & 3.0 &$  1.44\pm$0.04 & 3.8$\pm$2.3 & 4.5$\pm$2.5 & 4.9$\pm$1.7 & 317.0$\pm$16.3 \\
 9 & J1600$-$3053 & 244.35 &$ -10.07$& 106 &  8.1 & 1.0 &$  0.73\pm$0.32 & 1.2$\pm$0.6 & 5.8$\pm$4.3 & 2.9$\pm$0.4 & 320.4$\pm$15.3 \\
 10& J1614$-$2230 & 245.79 &$  -1.26$&  91 &  7.2 & 1.5 &$ -0.32\pm$0.04 & 1.2$\pm$0.8 &10.6$\pm$15.3& 5.1$\pm$0.7 & 324.3$\pm$15.1 \\
 11& J1640$+$2224 & 243.99 &$  44.06$& 110 & 11.1 & 0.5 &$  0.75\pm$0.02 & 0.6$\pm$0.7 & 6.2$\pm$4.9 &      -      &       -        \\
 12& J1643$-$1224 & 251.09 &$   9.78$& 122 & 11.2 & 3.1 &$-10.73\pm$0.15 & 4.6$\pm$2.5 & 6.1$\pm$6.7 & 5.7$\pm$1.7 & 232.9$\pm$16.8 \\
 13& J1713$+$0747 & 256.67 &$  30.70$& 209 & 10.9 & 0.3 &$ -0.14\pm$0.01 & 0.3$\pm$0.2 &11.2$\pm$6.6 & 1.3$\pm$0.1 & 327.7$\pm$15.2 \\
 14& J1738$+$0333 & 264.09 &$  26.88$&  54 &  6.1 & 3.0 &$ -7.75\pm$0.17 & 2.6$\pm$1.3 & 5.1$\pm$2.8 &      -      &       -        \\
 15& J1741$+$1351 & 264.36 &$  37.21$&  59 &  6.4 & 0.7 &$ -2.40\pm$0.05 & 1.1$\pm$0.9 & 5.0$\pm$2.9 & 0.6$\pm$0.5 &   0.6$\pm$17.4 \\
 16& J1744$-$1134 & 266.12 &$  11.81$& 116 & 11.4 & 0.7 &$ -0.17\pm$0.07 & 1.1$\pm$0.8 & 4.8$\pm$3.6 & 1.8$\pm$0.3 & 349.8$\pm$15.1 \\
 17& J1747$-$4036 & 267.58 &$ -17.20$&  54 &  3.8 & 6.0 &$-23.37\pm$1.49 &11.2$\pm$4.8 & 5.7$\pm$2.5 &      -      &      -         \\
 18& J1853$+$1303 & 286.26 &$  35.74$&  53 &  4.5 & 1.0 &$  0.77\pm$0.05 & 0.7$\pm$0.4 & 5.0$\pm$2.7 & 2.1$\pm$0.4 & 343.9$\pm$15.4 \\ 
 19& J1857$+$0943 & 286.86 &$  32.32$& 101 & 11.0 & 0.4 &$  3.01\pm$0.08 & 0.9$\pm$0.5 & 4.4$\pm$2.8 & 2.8$\pm$0.5 & 329.3$\pm$16.8 \\ 
 20& J1909$-$3744 & 284.22 &$ -15.16$& 165 & 11.2 & 0.5 &$ -2.08\pm$0.01 & 0.2$\pm$0.2 & 3.6$\pm$4.4 & 1.6$\pm$0.2 &   8.5$\pm$15.1 \\
 21& J1918$-$0642 & 290.31 &$  15.35$& 117 & 11.2 & 1.0 &$ -2.84\pm$0.11 & 0.9$\pm$0.5 & 4.7$\pm$2.5 & 2.1$\pm$0.4 &  16.4$\pm$15.4 \\
 22& J1923$+$2515 & 297.98 &$  46.70$&  48 &  4.3 & 0.3 &$  1.13\pm$0.12 & 1.6$\pm$1.7 & 4.9$\pm$2.5 &     -       &      -         \\
 23& J1939$+$2134 & 301.97 &$  42.30$& 165 & 11.3 & 1.0 &$ -0.10\pm$0.12 & 0.9$\pm$0.5 & 7.7$\pm$4.7 & 2.8$\pm$0.8 & 220.7$\pm$15.1 \\
 24& J1944$+$0907 & 300.00 &$  29.89$&  53 &  4.4 & 1.0 &$ -8.46\pm$1.12 & 1.2$\pm$1.1 & 4.4$\pm$1.7 &      -      &      -         \\ 
 25& J2010$-$1323 & 301.92 &$   6.49$&  88 &  6.2 & 0.9 &$  1.97\pm$0.15 & 1.7$\pm$1.1 & 6.7$\pm$10.1& 5.4$\pm$0.9 &  10.3$\pm$15.1 \\
 26& J2043$+$1711 & 318.87 &$  33.96$&  64 &  4.5 & 0.4 &$  0.61\pm$0.05 & 0.3$\pm$0.2 & 9.8$\pm$6.1 & 1.5$\pm$0.2 &  43.2$\pm$15.2 \\ 
 27& J2145$-$0750 & 326.02 &$   5.31$& 107 & 11.3 & 1.5 &$  1.19\pm$0.04 & 2.3$\pm$1.7 & 7.0$\pm$11.2& 3.0$\pm$0.8 &  58.7$\pm$17.0 \\
 28& J2214$+$3000 & 348.81 &$  37.71$&  53 &  4.2 & 3.0 &$  0.29\pm$0.10 & 7.9$\pm$5.7 & 6.1$\pm$3.0 &      -      &     -          \\
 29& J2302$+$4442 &   9.78 &$  45.67$&  58 &  3.8 & 2.5 &$ -4.77\pm$0.09 & 1.6$\pm$0.9 & 6.5$\pm$2.9 &      -      &     -          \\ 
 30& J2317$+$1439 & 356.13 &$  17.68$& 111 & 11.0 & 0.5 &$ -5.00\pm$0.06 & 0.3$\pm$0.2 & 6.2$\pm$5.2 & 2.2$\pm$0.3 &  76.6$\pm$15.1 \\
    \hline
    \multicolumn{12}{l}{\parbox[l]{170mm}{
        Notes for columns:
        column (1): index number;
        column (2): pulsar name;
        column (3) and (4): ecliptic longitude and latitude, in units of degree;
        column (5): number of observations for DM;
        column (6): span for DM data, in units of year;
        column (7): uncertainty median of DM measurements, in units of $\rm 10^{-4} cm^{-3} pc$;
        column (8): the average gradient for DM changes, obtained by a linear fitting to the trend curves, in units of $\rm 10^{-4} cm^{-3} pc$ per year;
        column (9): the median of instantaneous amplitudes for the noise term of DM changes,  in units of $\rm 10^{-4} cm^{-3} pc$;
        column (10): the median of instantaneous frequencies for the noise term of DM changes,  in units of $\rm 1/yr$;
        column (11): the amplitude of annual DM variations, in units of $\rm 10^{-4} cm^{-3} pc$;
        column (12): the peak phase of annual DM variations, referring to the beginning of a year, in units of degree (1 year = 360$^{\circ}$).
        }}
  \end{tabular}
\end{table*}